# Building and Installing a Hadoop/MapReduce Cluster from Commodity Components

Jochen L. Leidner / Gary Berosik

Technical Report

# Building and Installing a Hadoop/MapReduce Cluster from Commodity Components

by Jochen L. Leidner[1] and Gary Berosik[2]

**Abstract**
**This tutorial presents a recipe for the construction of a compute cluster for processing large volumes of data, using cheap, easily available personal computer hardware (Intel/AMD based PCs) and freely available open source software (Ubuntu Linux, Apache Hadoop).**

## Introduction

This article describes a straightforward way to build, install and operate a compute cluster from commodity hardware. A compute cluster is a *utility* that allows you to perform larger-scale computations faster than with individual PCs. We use commodity components (called "nodes") to keep the price down and to ensure easy availability of initial setup and replacement parts, and we use Apache Hadoop as a middleware for distributed data storage and parallel computing.

## Background

At the time of writing, single desktop computers and even mobile devices have become faster than the supercomputers of the past. At the same time, storage capacities of disk drives have been increasing by multiple orders of magnitude. As a result of mass production, prices have decreased and the number of users of such commodity machines has increased. At the same time, pervasive networking has become available and has led to the distribution and sharing of data, leading to distributed communication, creation, consumption, and collaboration. Perhaps paradoxically, the ever-increasing amount of digital content that is the result of more powerful machine storage and networking is leading to ever-increasing processing demands to find information and make sense of activities, preferences, and trends. The analysis of large networks such as the World Wide Web (WWW) is such a daunting task that it can only be carried out on a network of machines.

[1] Jochen L. Leidner, Ph.D. is a Research Scientist in the corporate Research and Development group at Thomson Reuters and a Director at Linguit Ltd. He holds a doctorate degree in Informatics from the University of Edinburgh, where he has been a Royal Society Enterprise Fellow in Electronic Markets and postdoctoral Research Fellow, and two Master's degrees in Computational Linguistics and English Language and Literature, and Computer Speech, respectively. His research interests include natural language processing, search engines, statistical data mining and software engineering. Jochen is a member of ACM, ACL and SIGIR, co-authored over twenty peer-reviewed papers and several patent applications.

[2] Gary Berosik is a Lead Software Engineer at Thomson Reuters Research and Development and an Adjunct Faculty member in the Graduate Programs in Software at the University of St. Thomas in St. Paul, MN. His interests include software engineering, parallel/grid/cloud processing, statistical machine learning algorithms, learning-support technologies, agent-based architectures and technologies supporting business intelligence and information analytics.

In the 1990s, Larry Page, Sergey Brin and others at Stanford University used a large number of commodity machines in a research project that attempted to crawl a copy of the entire WWW and analyze its content and hyperlink graph structure. The Web quickly grew, becoming too large for human-edited directories (e.g. Yahoo) to efficiently and effectively point people at the information they were looking for.  In response, Digital Equipment Corporation (DEC) proposed the creation of a keyword index of all Web pages, motivated by their desire to show the power of their 64-bit Alpha processor. This effort became known as the AltaVista search engine. Later, the aforementioned Stanford group developed a more sophisticated search engine named BackRub, later renamed Google. Today, Google is a search and advertising company, but is able to deliver its innovative services only due to massive investments in the large-scale distributed storage and processing capability developed in-house. This capability is provided by a large number of commodity off-the-shelf (COTS) PCs, the Google File System (GFS), a redundant cluster file system, and MapReduce -  parallel data processing middleware. More recently, the Apache Hadoop project has developed a reimplementation of parts of GFS and MapReduce, and many groups have subsequently embraced this technology, permitting them to do thing that they could not do on single machines.

## Procurement

We choose the GNU/Linux operating system because it is very efficient, scalable, stable, secure, is available in source code without licensing impediments, and has a large user base, which ensures rapid responses to support questions. We select the Ubuntu distribution of the operating system because it has good support, a convenient package management system, and is offered in a server edition that contains only server essentials (Ubuntu Server). At the time of writing, release 9.04 was current.  Little, if anything, of the described approach depends on this particular version.

You will need:

- $10,000 for 20 nodes; or, more precisely $500 + ($500 * n), where n is the number of nodes;
- an existing PC running Linux (or ready to be set up to run Linux, either as dual boot, in a virtualized setup, or exclusively), or $3,000 to buy one
- a screwdriver;
- a location to put the cluster, which has access to a power outlet and an Internet access point; and
- two person days (one to prepare, research, order, and another one to put things together once all components are at-hand).

We will need the following components:

- **Master**: We need a desktop PC running Linux as a master machine. In this project, no master purchase was necessary, because an existing DELL Optiplex (2x1 TB HDD configured as RAID, 24" TFT display) was already available.
  Cost: $0 (if you don't have a machine, I'd suggest ordering PowerMac or a DELL for $3,000).
- **Switch**: Using a professional-grade Netgear ProSafe 24 means we can save the time setting up the machine (this switch detects everything automatically once it is plugged in). It has 24 sockets, which means we can have up to 22 nodes (we have to connect the master and the network cable to two of the sockets).
- **Network cabling**: Get at least one CAT6 Ethernet patch cable per node to connect the node with the switch. The cables should be as short as possible but long enough to bridge the distance between switch and back of the node's case. If you can, buy half a dozen of these extra. The price should be around $3 if ordered online, or $20 in a store. Cost: $3*6 + ($3 * n), plan $80 for a 20-node cluster.
- **Nodes**: We should plan for at least two slave node PCs in order to have an advantage over a single, powerful desktop PC or server and to deploy the various core Hadoop processes. Since we are drawing from commodity components, we should pick an attractive package deal rather than waste time on customizing a node. Criteria are: price, CPU speed, RAM, number of hard disk slots, energy consumption, cooling and noise. The hard disk drive size is *not* a criterion, because they will be replaced (cheap commodity PC deals include very small disk drives). The amount of RAM is important, but since it can be cheaply replaced what matters is more the maximum amount possible rather than the memory size it comes from the factory with. 2 GB should be considered the absolute minimum, 4 GB/node RAM is recommended (16 GB would be ideal, but at the time of writing is unlikely to be found in consumer machines).
  For this project, Acer X2 with a dual-core Athlon X2 64-bit CPU (1.2 GHz), 3 GB RAM, 320 GB HDD, was selected, because it offers 4,800 bogomips/core for under $400. The disk drives were replaced due to their insufficient size. Cost: $400 * n, plan $10,000 for 20 nodes (if you buy the nodes incrementally over time, you can expect either the price to go down, or to get more powerful machines; the disadvantage is higher support cost in time, because you will have a whole zoo of different pieces of hardware.)
- **Hard disk drives** (1 TB or higher). Determine the type of hard disk drive with the lowest $/TB cost. High-capacity drives are overly expensive, and the average size of drives used in commodity PCs as you buy them is too small, so the drives they come with (e.g. 320 GB) have to be replaced. Cost: $89 * n, plan $1,780 for 20 nodes.
- **Enclosure** (shelf or rack). The nodes, switch etc. need to live somewhere. A 42U 19" rack is the standard for data centers, however it may prove an unreasonable choice for several reasons: first, the cost of a new rack could easily exceed the total of the hardware expenses for the cluster itself, and secondly, since the nodes are commodity machines as opposed to

"professional" 19" servers, they may be hard to fix, so the main advantage of the 19" rack may be lost on them.

Alternatives are cheap IKEA shelves made from wood or metal, or anything similar. One valuable feature for situations where the height of the rack does not exceed its width/depth is to attach some wheels to keep it mobile (once the nodes are put on it, it can get quite heavy, and we would like to avoid having to disassemble the cluster if the need for a relocation arises). Cost: approximately $400 (but it is likely that you may already have a spare shelf somewhere).

- **Socket multiplier**. Use a socket multiplier to cope with the plugs of all nodes, the master, the switch and other devices you may have. Make sure the socket multipliers are fused and the power outlet itself can actually sustain the power drawn without blowing the fuse of the room/building where the cluster lives. Cost: $20.
- Optionally, you can use an **Uninterruptible Power Supply** (UPS) unit if your computations run for more than a few days (e.g. http://www.tripplite.com/ ), but the project as described and implemented here did not exercise that option.

## Physical Setup & Assembly

It is important to pick a suitable location for the cluster, which is best chosen upfront. Cable length planning is important. The location must be close to the Internet access point and should be protected from unauthorized or accidental access, and must be well ventilated.

Power consumption for a cluster of reasonable size will be considerable. For instance, ten nodes of Acer X1700 at 220 Watts each amount to 2200/120 = 18.4 Amperes, which is just under the limit of what the circuit breaker of a typical household or small office will be able to carry. In addition, consider the significant heat generation that a large cluster can create.

Open the node PCs and remove their internal HDDs. Now is a good time to plug in additional RAM before inserting the new 1 TB replacement hard disk drives. After replacing the first disk, boot the Ubuntu DVDROM *while the case is still open* in order to ensure that the disks are recognized by the BIOS and by Ubuntu (i.e. skip down to the next Section, Operating System Installation for one node). Once you know that your disks are compatible with your nodes' SATA controllers and your BIOS, close all nodes' cases.

If using a closed 19" rack that comes with a door (recommended to keep the noise level down, but even more expensive than its open siblings), you may consider removing all cases of the node PCs altogether to improve ventilation (this "naked" configuration was pioneered by Google).

## Operating System Installation

First make sure you are connected to the Internet. You can either download Ubuntu Server 9.07 or higher and burn a DVD with this ISO image, or purchase a

DVDROM, but in the latter case you will still require Internet access in order to obtain security updates and install additional software.

Connect all nodes with the switch using the CAT6 cables and connect all nodes with the power source using the socket multiplier. Connect the master PC's screen to the first node to be installed, insert the DVDROM with the operating system and switch it on to boot Ubuntu Server. Select "Install Ubuntu Server" (first menu option). Select the language, keyboard layout (e.g. English, U.S.) using the keyboard (confirm with RETURN, switch boxes on the screen with TAB) and set up the partitions as follows:

- **boot** (1 GB) not mounted, ext3, bootable, primary (important: this one should be at the beginning of the disk, or you may run into BIOS boot problems such as "grub error 2" etc.);
- **root** (50 GB) mounted as /, ext3, logical;
- **main** (0.94 TB) mounted on /var (var is a standard convention indicating variable data, i.e. a high amount of input/output is to be expected), ext3, logical; and
- **swap** (2 GB) not mounted, type swap, logical using manual partitioning (last option). Don't use logical volume management (LVM) or encryption.

After saving/applying these settings, the node will be busy for quite some time (do not worry if the progress bar moves slowly or stalls for significant amount of time, 1 TB is a lot of formatting to do!).

Set the time zone to UTC/GMT rather that a specific location, especially if your cluster is distributed over multiple geographies; the rationale here is that log files with time stamps are easier to read if all servers run in the same time zone.

Next, select and install software packages:

- Ubuntu server (always included, no action required);
- LAMP;
- OpenSSH; and
- Samba (optional; if you may want to use nodes as file servers outside of the Hadoop file system, for example if you use the cluster at home)

You will also have to give the computer a name (I named my cluster hydra, so the nodes are called hydra1, hydra2, ...). It then attempts to detect the network and obtains an IP address via DHCP. You will be asked to define a non-root account with a password as well.

During package installation, do not set any database passwords (leave the respective fields empty and confirm) to keep life simple. If your cluster is part of a larger infrastructure or exposed to an external network, consider using the master PC as a firewall by installing a second network card that can become the sole gateway to the external network (in this case, make sure all services other than HTTP (port 80) and ssh (port 23) are disabled).

After the installation of the core packages is completed, the DVDROM is ejected and the system will re-boot, hopefully successfully from the new drive for the first time.

Log on using your new account and perform the following operations to ensure we have the latest stable software states:

```
sudo apt-get update
sudo apt-get dist-upgrade
```

We also want to be able to use the X11 networking system on each node, in order to make the administration easier (graphical login) and to run programs on particular nodes manually while directing their output to the master. So let's get:

```
sudo apt-get install xorg gdm xfce4
```

For added convenience, we want XEmacs on each node:

```
sudo apt-get install xemacs21
```

Of course we should only do this on an experimental/research cluster; we won't need (and should not install) X11 or XEmacs on a production server in 99.99% of cases.

Depending on the tasks that we anticipate using the cluster for, we may want to consider installing additional packages. I do a lot of Web research, data crawling, analytics and statistical machine learning, so it makes sense to get some crawlers, the R statistics system and libraries for numeric computing:

```
sudo apt-get install wget curl lynx

sudo apt-get install r-base r-base-dev

sudo apt-get install python-numpy python-scipy
```

For Hadoop cluster experiments, we'll need a recent Java Development Kit (JDK):

```
sudo apt-get install sun-java6-jdk
```

You will be asked to accept Java's license terms.

At this point, our first node is operational as far as the operating system is concerned, but in order to make combined used of its nodes as a single quasi-utility, we still need to install Apache Hadoop, and then replicate the setup to the other nodes. If you are not using Hadoop, but another parallel computing middleware (for example a batch system like Condor or Sun GridEngine), you can skip the next part.

## Example Hadoop Installation on a Small Cluster

After the installation of the operating system, we can now turn to the setup of Apache Hadoop as our operating system middleware for redundant storage and parallel processing. As an example, we describe a three-node installation.

---

util-egn-1      DELL Optiplex GX260        Master
        NameNode;DataNode;TaskTracker

util-egn-2    DELL Optiplex 745    Slave
        SecondaryNameNode;DataNode;TaskTracker

util-egn-3    DELL Optiplex 755    Slave   JobTracker;DataNode;TaskTracker

---

1. Insure Java is set up. The latest Hadoop releases depend on a Java 6.x or later release version. Download/install/test the latest 6.x Java release if this is not already set up.

It is recommended to follow the procedures for single node cluster setup as described in the online article by Michael G. Noll on running Hadoop in Ubuntu Linux environments (1).

(Per this article's guidance, do this on each machine of a multi-node cluster to help verify the operational status of Hadoop on each node before continuing to set up a multi-node configuration.)

The following steps show examples of following these instructions.

---

2. Add a hadoop group and hadoop user for that group.

```
<your-user-name>@util-egn-1:~$ sudo addgroup hadoop
<your-user-name>@util-egn-1:~$ sudo adduser --ingroup hadoop hadoop
```

As needed, switch user to hadoop:

```
su - hadoop
```

---

3. Add the following exports for proxy and JAVA_HOME to the .bashrc file for both your user and the new hadoop user.

```
export http_proxy=<yourProxy>:<proxyPort>
export JAVA_HOME=<yourJavaHomePath>
```

---

4. Configure and test SSH operation on all nodes (required by Hadoop).

As the hadoop user, on the master node, create the RSA key:

```
hadoop@util-egn-1:~$ ssh-keygen -t rsa -P ""
```

Copy or append the new key to the authorized_keys file in the .ssh directory.

```
hadoop@util-egn-1:~$ cat /home/hadoop/.ssh/id_rsa.pub >>
~hadoop/.ssh/authorized_keys
```

Try to connect to the local machine with the hadoop user.  Respond with a **yes** when prompted to "continue connecting".

```
hadoop@util-egn-1:~$ ssh localhost
```

5. Download Hadoop 0.19.2

As of this writing, there are known instability issues with the 0.20 release, so release 0.19.2 is used for this installation.

For example, download from:

```
http://newverhost.com/pub/hadoop/core/hadoop-0.19.2/hadoop-
0.19.2.tar.gz
```

Now, with administrator permissions, install this release in the desired directory location. (The example installation steps below assume the original download was to the directory location: /home/<your-user-name>/Desktop.)

Note that *these steps must be performed with administrator level permissions*.

Uncompress the hadoop release to the desired location:

```
root@util-egn-1:/usr/local# sudo tar xzf /home/<your-user-
name>/Desktop/hadoop-0.19.2.tar.gz
```

Rename the release as desired, and change ownership of all the release contents to permit use by the hadoop user.

```
root@util-egn-1:/usr/local# sudo mv hadoop-0.19.2 hadoop
root@util-egn-1:/usr/local# sudo chown -R hadoop:hadoop hadoop
```

6. Add an export for HADOOP_HOME to the .bashrc file for both your user and the hadoop user.

```
export HADOOP_HOME=<yourHadoopHomePath>
```

7. Edit the file: /usr/local/hadoop/conf/hadoop-env.sh by uncommenting the export for JAVA_HOME and setting it to the correct value.

```
export JAVA_HOME=<yourJavaHomePath>
```

8. Edit the file <yourHadoopHomePath>/conf/hadoop-site.xml to contain single node test configuration settings. Adjust values to suit your own configuration needs. There are MANY possible default configuration parameter settings that can be adjusted. See the <yourHadoopHomePath>/conf/hadoop-defaults.xml file for more information about the complete set of adjustable parameters.

```
<configuration>

<property>
  <name>hadoop.tmp.dir</name>
  <value>/usr/local/hadoop/tmp/datastore/hadoop-${user.name}</value>
  <description>
```

```
  A base location for other temp datastore directories.
  </description>
</property>

<property>
  <name>fs.default.name</name>
  <value>hdfs://localhost:54310</value>
  <description>
  The name of the default file system. A URI whose
  scheme and authority determine the FileSystem implementation. The
  URI's scheme determines the config property (fs.SCHEME.impl) naming
  the FileSystem implementation class. The URI's authority is used to
  determine the host, port, etc. for a FileSystem.
  </description>
</property>

<property>
  <name>mapred.job.tracker</name>
  <value>localhost:54311</value>
  <description>
  The host and port that the MapReduce job tracker runs at.
  If "local", then jobs are run in-process as a single map and reduce
task.
  </description>
</property>

<property>
  <name>dfs.replication</name>
  <value>1</value>
  <description>
  Default block replication.
  The actual number of replications can be specified when the file is
created.
  The default is used if replication is not specified at create time.
  </description>
</property>

</configuration>
```

9. With administrator privileges, create the hadoop temp datastore directory for your user and the hadoop user and change ownership to allow use by the hadoop and your user:

```
root@util-egn-1:/usr/local/hadoop# mkdir tmp
root@util-egn-1:/usr/local/hadoop# mkdir tmp/datastore
root@util-egn-1:/usr/local/hadoop# mkdir tmp/datastore/hadoop-
hadooproot@util-egn-1:/usr/local/hadoop# sudo chown -R hadoop:hadoop
  tmp/datastore/hadoop-hadoop

root@util-egn-1:/usr/local/hadoop# mkdir tmp/datastore/hadoop-<your-
user-name>
root@util-egn-1:/usr/local/hadoop# sudo chown -R <your-user-
name>:<your-user-name>
  tmp/datastore/hadoop-<your-user-name>
```

Verify proper creation and ownership of the temp datastore directories:

```
drwxr-xr-x 4 root root 4096 2009-10-22 12:11 datastore
drwxr-xr-x 2 hadoop  hadoop  4096 2009-10-22 12:09 hadoop-hadoop
```

## 10. Test Hadoop execution in single-node mode, using HDFS.

As the hadoop user, format the NameNode.

```
hadoop@util-egn-1:~$ $HADOOP_HOME/bin/hadoop namenode -format
```

Start the (single-node) cluster.

```
hadoop@util-egn-1:~$ $HADOOP_HOME/bin/start-all.sh
```

Verify that the expected Hadoop processes are running using Java's jps.

```
hadoop@util-egn-1:~$ jps
```

27069 JobTracker

26641 NameNode

26729 DataNode

29425 Jps

26923 SecondaryNameNode

27259 TaskTracker

With administrator permission, use netstat to verify that Hadoop is listening on the expected/configured ports.  For example:

```
root@util-egn-1:~# sudo netstat -plten | grep java
```

```
tcp6  0  0 :::50020     :::*      LISTEN  1001   607378   26729/java
tcp6  0  0 :::45700     :::*      LISTEN  1001   591676   26729/java
tcp6  0  0 127.0.0.1:54310  :::*  LISTEN  1001    590635   26641/java
tcp6  0  0 127.0.0.1:54311  :::*  LISTEN  1001    594563   27069/java
tcp6  0  0 :::50090     :::*      LISTEN  1001   601284   26923/java
tcp6  0  0 :::50060     :::*      LISTEN  1001   601022   27259/java
tcp6  0  0 :::50030     :::*      LISTEN  1001   600371   27069/java
tcp6  0  0 :::41071     :::*      LISTEN  1001   592854   26923/java
tcp6  0  0 127.0.0.1:51633  :::*  LISTEN  1001    601104   27259/java
tcp6  0  0 :::50070     :::*      LISTEN  1001   595090   26641/java
tcp6  0  0 :::50010     :::*      LISTEN  1001   595317   26729/java
tcp6  0  0 :::38554     :::*      LISTEN  1001   593715   27069/java
tcp6  0  0 :::50075     :::*      LISTEN  1001   601394   26729/java
tcp6  0  0 :::39071     :::*      LISTEN  1001   590581   26641/java
```

## 11. Set up and run a Hadoop example test application to verify operability.

Adjust the <HadoopHome>/conf/hadoop-env.sh file to set JAVA_HOME, HADOOP_HOME, and a reasonable CLASSPATH (if desired) for Hadoop executions.

```
export JAVA_HOME=/usr/lib/jvm/java-6-sun/jre
export HADOOP_HOME=/usr/local/hadoop
```

```
export CLASSPATH=$HADOOP_HOME/hadoop-0.19.2-
core.jar:$HADOOP_HOME/hadoop-0.19.2-examples.jar:$HADOOP_HOME/hadoop-
0.19.2-test.jar:$HADOOP_HOME/hadoop-0.19.2-
tools.jar:$CLASSPATH:$classpath
```

Run the test example program.

(The following example executes the pi program included in the distributed Hadoop examples. Use the *source* command to insure that the settings are kept by the executing shell process.)

```
hadoop@util-egn-1:~$ source sethadenv.sh

hadoop@util-egn-1:~$ $HADOOP_HOME/bin/hadoop jar $HADOOP_HOME/hadoop-
0.19.2-examples.jar pi 2 10
```

The output should look similar to the following:

```
Number of Maps = 2 Samples per Map = 10
Wrote input for Map #0
Wrote input for Map #1
Starting Job
09/10/22 13:17:50 INFO mapred.FileInputFormat: Total input paths to
process : 2
09/10/22 13:17:50 INFO mapred.JobClient: Running job:
job_200910221225_0001
09/10/22 13:17:51 INFO mapred.JobClient:  map 0% reduce 0%
09/10/22 13:18:00 INFO mapred.JobClient:  map 50% reduce 0%
09/10/22 13:18:03 INFO mapred.JobClient:  map 100% reduce 0%
09/10/22 13:18:10 INFO mapred.JobClient:  map 100% reduce 100%
09/10/22 13:18:11 INFO mapred.JobClient: Job complete:
job_200910221225_0001
09/10/22 13:18:11 INFO mapred.JobClient: Counters: 16
09/10/22 13:18:11 INFO mapred.JobClient:   File Systems
09/10/22 13:18:11 INFO mapred.JobClient:     HDFS bytes read=236
09/10/22 13:18:11 INFO mapred.JobClient:     HDFS bytes written=212
09/10/22 13:18:11 INFO mapred.JobClient:     Local bytes read=78
09/10/22 13:18:11 INFO mapred.JobClient:     Local bytes written=218
09/10/22 13:18:11 INFO mapred.JobClient:   Job Counters
09/10/22 13:18:11 INFO mapred.JobClient:     Launched reduce tasks=1
09/10/22 13:18:11 INFO mapred.JobClient:     Launched map tasks=2
09/10/22 13:18:11 INFO mapred.JobClient:     Data-local map tasks=2
09/10/22 13:18:11 INFO mapred.JobClient:   Map-Reduce Framework
09/10/22 13:18:11 INFO mapred.JobClient:     Reduce input groups=2
09/10/22 13:18:11 INFO mapred.JobClient:     Combine output records=0
09/10/22 13:18:11 INFO mapred.JobClient:     Map input records=2
09/10/22 13:18:11 INFO mapred.JobClient:     Reduce output records=0
09/10/22 13:18:11 INFO mapred.JobClient:     Map output bytes=64
09/10/22 13:18:11 INFO mapred.JobClient:     Map input bytes=48
09/10/22 13:18:11 INFO mapred.JobClient:     Combine input records=0
09/10/22 13:18:11 INFO mapred.JobClient:     Map output records=4
09/10/22 13:18:11 INFO mapred.JobClient:     Reduce input records=4
Job Finished in 21.342 seconds
Estimated value of PI is 3.2

hadoop@util-egn-1:~$
```

Congratulations! At this point you have a simple, single-node Hadoop environment up and running!

12. Shut down the Hadoop processes in the single-node cluster.

```
hadoop@util-egn-1:~$ $HADOOP_HOME/bin/stop-all.sh
```

The output should look similar to the following:

```
stopping jobtracker
localhost: stopping tasktracker
stopping namenode
localhost: stopping datanode
localhost: stopping secondarynamenode
```

To configure the Hadoop middleware to handle a *multi*-node cluster, we recommend you follow the procedures for setting up multi-node clusters described in the article by Michael G. Noll (2).

You now face the issue of having to install the whole node's environment (Linux, packages, Hadoop) from one node to the remaining nodes in a near-identical way. For smaller clusters this can be done manually. For larger clusters with nodes that have possibly different hardware specifications, stronger tools need to be used to define machine classes, separate configurations for each class, and assist in the distribution of these configurations to the appropriate node machines. In these settings, various sources suggest the use of configuration management tools like Puppet (4), cfengine (5), or bcfg2 (6).

More concretely, there are several solutions to this, depending on your experience and number of nodes:

(a) burn an ISO image with your setup and use this with the remaining nodes;
(b) insert the empty hard disk drives as secondary drives in the master PC temporarily in order to copy over the entire disk using the **dd(1)** command;
(c) clone the disk over a network connection using **dd(1)** and netcat (**nc(1)**) as outlined by (7); or
(d) install the other nodes manually (estimated time: about 30 min/node).

Method (c) is superior for large clusters, but (d) is fast enough for smaller clusters. Remember that the hostname must be unique, so you may have to set it manually after cloning the node setups by manually invoking the **hostname(1)** command for each node.

In order to automate the installation completely, (9) recommends using static IP addresses for the nodes, setting the hostname by keeping a file `hostnames.new` that contains the node names and their static IP addresses, and then generating a set of node-specific kick-start files from a master template (here called "anaconda-ks.cfg", with NODE_HOSTNAME and NODE_STATIC_IP being placeholders) as follows:

```
for i in $(cat ~/hostnames.new) ; do \
   cat anaconda-ks.cfg | sed s/NODE_HOSTNAME/$i/g | sed
s/NODE_STATIC_IP/$(grep $i /etc/hosts | awk '{print
$1}')/g > ks-$i.cfg ; \
done
```

In their approach, the operating system is booted over the network using the node-specific kick-start file.

But there is yet another mode of operation to install Hadoop on more than one node very conveniently: Cloudera Inc. (a cluster/cloud computing startup, which recently hired Hadoop architect Doug Cutting) permits you to enter your desired cluster configuration on a Web interface (my.cloudera.com), and it automatically creates customized installers (e.g. *.rpm packages) that contain all the cluster configuration information.

Now that your Hadoop cluster is fully operational, we recommend you try out the word count example from the Apache Hadoop tutorial (8), which shows you how the UNIX **wc(1)** command can be distributed across a cluster.

## Summary and Conclusion

We have described a successfully completed project to build a cluster computing utility from commodity parts. The cluster is affordable (<$10,000), can be built incrementally, and is more powerful than servers that were priced over a quarter million dollars just a few years ago. Hadoop provides powerful operating system middleware for large-scale batch processing such as the automatic analysis of large data collections. We expect that in the future, enterprise versions of commodity operating systems will incorporate some of these capabilities, but hope the above introduction can serve to give the interested reader a head start.

*Happy Hadooping!*

6. **bcfg2** online information:
   http://trac.mcs.anl.gov/projects/bcfg2/wiki/UsingBcfg2
7. Web article by Vivek Gite, **Copy hard disk or partition image to another system using a network and netcat (nc).**
   http://www.cyberciti.biz/tips/howto-copy-compressed-drive-image-over-network.html
8. **Apache Hadoop Map/Reduce Tutorial** online information:
   http://hadoop.apache.org/common/docs/current/mapred_tutorial.html
9. **Installing CentOS On a Cluster Via NFS** online information:
   http://biowiki.org/InstallingCentOSOnClusterViaNFS

**Disclaimer.**
All opinions expressed in this article are the authors', and do not reflect any official opinion or endorsement by the Thomson Reuters Corporation.

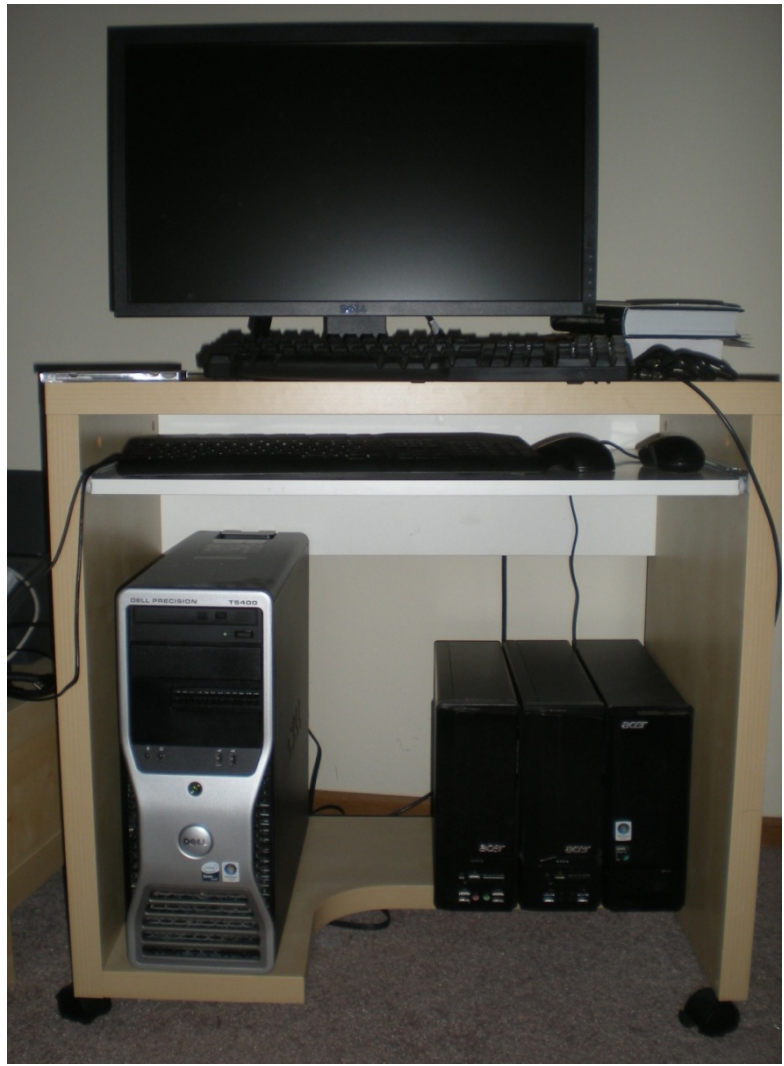

**Figure 1.** Hydra, a miniature cluster with one master PC and three nodes.